\documentclass[11pt,twocolumn]{article}

\usepackage[utf8]{inputenc}
\usepackage[T1]{fontenc}
\usepackage{times}
\usepackage{graphicx}
\usepackage{booktabs}
\usepackage{amsmath}
\usepackage{url}
\usepackage[margin=1in]{geometry}
\usepackage{caption}
\usepackage{multirow}
\usepackage{xcolor}
\usepackage{enumitem}
\usepackage{placeins} 
\usepackage{subcaption}
\usepackage{hyperref}
\setlist{nosep}

\title{Decomposing Docker Container Startup Performance:\\A Three-Tier Measurement Study on Heterogeneous Infrastructure}

\author{
    Shamsher Khan \\
    \textit{Senior DevOps Engineer} \\
    \textit{GlobalLogic (Hitachi Group)} \\
    \textit{IEEE Senior Member} \\
    Email: \protect\href{shamsher.khan.research@gmail.com}{shamsher.khan.research@gmail.com}
}
\date{February 2026}

\begin{document}
\maketitle

\begin{abstract}
Container startup latency is a critical performance metric for CI/CD pipelines, serverless computing, and auto-scaling systems, yet practitioners lack empirical guidance on how infrastructure choices affect this latency. We present a systematic measurement study that decomposes Docker container startup into constituent operations across three heterogeneous infrastructure tiers: Azure Premium SSD (cloud SSD), Azure Standard HDD (cloud HDD), and macOS Docker Desktop (developer workstation with hypervisor-based virtualization). Using a reproducible benchmark suite that executes 50 iterations per test across 10 performance dimensions, we quantify previously unmeasured relationships between infrastructure configuration and container runtime behavior. Our key findings include: (1)~container startup is dominated by runtime overhead rather than image size, with only 2.5\% startup variation across images ranging from 5\,MB to 155\,MB on SSD; (2)~storage tier selection imposes a $2.04\times$ startup penalty (HDD 1157\,ms vs.\ SSD 568\,ms); (3)~Docker Desktop's hypervisor layer introduces a $2.69\times$ startup penalty and $9.5\times$ higher CPU throttling variance compared to native Linux; (4)~OverlayFS write performance collapses by up to two orders of magnitude compared to volume mounts on SSD-backed storage; and (5)~Linux namespace creation contributes only 8--10 \,ms ($<1.5\%$) of total startup time. All measurement scripts, raw data, and analysis tools are publicly available.
\end{abstract}


\section{Introduction}
\label{sec:intro}

Linux containers have become the dominant deployment unit for native cloud applications, with Docker serving as the de facto standard runtime for development and CI/CD environments~\cite{casalicchio2019,vayghan2021}. The performance characteristics of containers, particularly the startup latency, directly impact developer productivity, CI/CD pipeline throughput, and auto-scaling responsiveness~\cite{shahrad2020,jonas2019}.

Despite widespread adoption, there is a persistent gap between development-environment performance and production behavior. Practitioners commonly observe that containers behave differently across infrastructure tiers, yet empirical measurements quantifying these differences are largely absent from the literature. Existing research has focused primarily on comparing containers with virtual machines~\cite{felter2015,seo2014,gupta2025} or optimizing container startup for serverless workloads~\cite{oakes2018,du2020,akkus2018}, treating the container runtime as a monolithic unit rather than decomposing it into individually measurable components.

This paper addresses three research questions:

\noindent\textbf{RQ1 (Decomposition):} How does container startup time decompose into constituent operations, and which operations are infrastructure-dependent versus architecture-invariant?

\noindent\textbf{RQ2 (Invariants):} Which container runtime primitives exhibit platform-invariant behavior, and what is the practical significance of this invariance for infrastructure planning?

\noindent\textbf{RQ3 (Storage Tier Impact):} Under what conditions does storage tier selection materially impact container performance, and what are the quantifiable trade-offs?

\subsection{Contributions}

This paper makes the following contributions:

(1)~A systematic decomposition of container startup latency into kernel-level primitives (namespace creation, cgroup configuration) and infrastructure-dependent operations (OverlayFS mount, storage I/O), measured across three heterogeneous tiers with statistical rigor ($n\!=\!50$ iterations, 95\% confidence intervals, non-parametric significance tests).

(2)~Empirical quantification of the Docker Desktop virtualization tax: a $2.69\times$ startup penalty and $9.5\times$ higher CPU throttling variance compared to native Linux.

(3)~Identification of a storage-layer performance paradox in OverlayFS: write throughput drops to $0.11\text{--}0.15\times$ of volume mount speed, while metadata operations are $1.3\text{--}4.8\times$ faster on OverlayFS than on volume mounts.

(4)~An open-source benchmark suite that enables reproducible cross-platform container performance evaluation~\cite{khan2025toolkit}.

\subsection{Paper Organization}

Section~\ref{sec:related} discusses related work. Section~\ref{sec:method} describes our methodology. Section~\ref{sec:results} presents results organized by research question. Section~\ref{sec:discussion} discusses implications. Section~\ref{sec:limitations} addresses limitations. Section~\ref{sec:conclusion} concludes.

\section{Related Work}
\label{sec:related}

\subsection{Container vs.\ Virtual Machine Performance}

The foundational work by Felter et al.~\cite{felter2015} provided the first rigorous comparison between Docker containers and KVM virtual machines, demonstrating near-native container performance with overhead concentrated in I/O-intensive workloads using the AUFS storage driver. Seo et al.~\cite{seo2014} confirmed faster startup for containers versus VMs. Casalicchio~\cite{casalicchio2019} surveyed container orchestration, noting that the startup latency varies with the orchestration overhead. Ferreira and Sinnott~\cite{ferreira2019} found bounded orchestration overhead on managed Kubernetes. Gupta and Nahrstedt~\cite{gupta2025} characterized containers on edge devices, reporting minimal CPU overhead but significant storage degradation.

These studies treat the container runtime as a monolithic unit. Our work differs by decomposing the startup into individually measurable components.

\subsection{Container Startup Optimization}

Oakes et al.~\cite{oakes2018} analyzed Linux container primitives and implemented SOCK for serverless workloads, confirming that kernel-level isolation contributes minimal latency. Du et al.~\cite{du2020} proposed a Catalyzer for sub-millisecond startup via snapshot-based initialization. Akkus et al.~\cite{akkus2018} presented SAND for serverless optimization. Grambow et al.~\cite{grambow2023} provided systematic benchmarking for container orchestrators.

Our work complements these efforts by quantifying the decomposition of startup latency across infrastructure tiers, enabling informed optimization decisions.

\subsection{Storage-Layer Performance}

Sharma et al.~\cite{sharma2016} studied containers at scale, finding that storage I/O contention compounds overhead. The Linux kernel documentation on OverlayFS~\cite{overlayfs} and cgroups v2~\cite{cgroups} provides an architectural context. Empirical measurements of copy-up overhead and write performance across storage tiers are largely absent from the literature, a gap that our work addresses.

\subsection{Measurement Methodology}

Gregg~\cite{gregg2020} established systematic approaches for Linux performance analysis. Ferraro Petrillo et al.~\cite{petrillo2024} analyzed container engines using 100+ iterations with confidence intervals--practices we adopt. Iosup et al.~\cite{iosup2011} demonstrated the importance of workload characterization.

\subsection{Positioning}

Our work uniquely combines: (1)~systematic decomposition of container startup, (2)~cross-tier comparison across heterogeneous storage, and (3)~identification of platform-invariant primitives alongside platform-dependent operations. Table~\ref{tab:positioning} summarizes.

\begin{table}[htbp]
\centering
\caption{Comparison with related work.}
\label{tab:positioning}
\footnotesize
\begin{tabular}{@{}lccc@{}}
\toprule
\textbf{Study} & \textbf{Scope} & \textbf{Dec.} & \textbf{Tiers} \\
\midrule
Felter~\cite{felter2015} & Single host & No & No \\
Oakes~\cite{oakes2018} & Serverless & Part. & No \\
Du~\cite{du2020} & Serverless & Yes & No \\
Petrillo~\cite{petrillo2024} & Multi-eng. & No & No \\
Gupta~\cite{gupta2025} & Edge/IoT & No & No \\
\textbf{This work} & \textbf{Multi-tier} & \textbf{Yes} & \textbf{Yes} \\
\bottomrule
\end{tabular}
\end{table}

\FloatBarrier

\section{Methodology}
\label{sec:method}

\subsection{Platform Configuration}

We selected three platforms that represent different infrastructure tiers (Table~\ref{tab:platforms}): Azure Premium SSD and Standard HDD share identical VM specifications (Standard\_D2s\_v3, 2~vCPU, 8 \, GB), isolating the storage tier as the independent variable. The Standard\_D2s\_v3 configuration represents the most commonly deployed general-purpose VM tier in Azure and is frequently used for CI/CD build agents and development workloads~\cite{shahrad2020}. The macOS platform introduces hypervisor-based virtualization via Docker Desktop's LinuxKit VM. We emphasize that the macOS results are intended to characterize developer environments rather than provide an architectural comparison with cloud CPUs.

\begin{table}[htbp]
\centering
\caption{Platform configurations.}
\label{tab:platforms}
\footnotesize
\begin{tabular}{@{}llll@{}}
\toprule
& \textbf{Prem.\ SSD} & \textbf{Std HDD} & \textbf{macOS DD} \\
\midrule
CPU & 2 vCPU (Xeon) & 2 vCPU (Xeon) & M1 Pro \\
Mem. & 8\,GB & 8\,GB & 16\,GB \\
Stor. & SSD (LRS) & HDD (LRS) & APFS/VM \\
Docker & 28.x & 28.x & 28.4.0 \\
Driver & overlay2 & overlay2 & overlayfs \\
Virt. & None & None & HV.f \\
\bottomrule
\end{tabular}
\end{table}

\subsection{Container Images}

Three images spanning 5--155 \, MB: \texttt{alpine:latest} (3 \, MB, 1~layers), \texttt{nginx:latest} (67 \, MB, 7~layers) and \texttt{python:3.11-slim} (155 \, MB, 5~layers).

\subsection{Measurement Procedure}

All measurements were automated using a bash script that recorded timing data in CSV format. Timing uses nanosecond-precision host-side timestamps. For warm starts, images were pre-pulled. For cold starts, the file system caches were cleared by \texttt{drop\_caches} with a 2-second sleep between iterations. On macOS, cache clearing cannot reach Docker Desktop's internal VM caches, documented as a limitation (Section~\ref{sec:limitations}).

\subsection{Statistical Methods}

Each measurement was repeated 50 times ($n\!=\!10$ for pull tests, $n\!=\!20$ for cold starts). We report the mean ($\mu$), standard deviation ($\sigma$), and 95\% confidence interval:
\begin{equation}
\text{CI}_{95} = \mu \pm 1.96 \times \frac{\sigma}{\sqrt{n}}
\end{equation}

We applied the Mann-Whitney U test for significance ($\alpha\!=\!0.05$) and Cliff's delta ($d$) for effect size: $|d| < 0.147$ (negligible), $< 0.33$ (small), $< 0.474$ (medium), $\geq 0.474$ (large). Raw data and scripts are available at~\cite{khan2025toolkit}.

\FloatBarrier

\section{Results}
\label{sec:results}

\subsection{Container Startup Latency (RQ1)}

\subsubsection{Warm-Start Performance}

Table~\ref{tab:warmstart} presents warm-start latency across all platform-image combinations.

\begin{table}[htbp]
\centering
\caption{Warm-start latency (ms). Mean $\pm$ 95\% CI ($\sigma$). $n\!=\!50$.}
\label{tab:warmstart}
\small
\begin{tabular}{@{}lccc@{}}
\toprule
\textbf{Platform} & \textbf{alpine} & \textbf{nginx} & \textbf{python} \\
\midrule
Prem.\ SSD & $568 \pm 5$ & $564 \pm 12$ & $554 \pm 6$ \\
 & ($\sigma\!=\!19$) & ($\sigma\!=\!42$) & ($\sigma\!=\!22$) \\
\addlinespace
Std HDD & $1157 \pm 27$ & $1287 \pm 36$ & $1334 \pm 35$ \\
 & ($\sigma\!=\!96$) & ($\sigma\!=\!131$) & ($\sigma\!=\!126$) \\
\addlinespace
macOS DD & $1528 \pm 139$ & $1850 \pm 155$ & $1859 \pm 139$ \\
 & ($\sigma\!=\!502$) & ($\sigma\!=\!560$) & ($\sigma\!=\!502$) \\
\bottomrule
\end{tabular}
\end{table}

\textbf{Finding 1: Image size has minimal impact on startup.} On Premium SSD, warm-start latency varies only 554--568\,ms across images ranging from 5\,MB to 155\,MB---a 2.5\% difference. Runtime overhead (namespace creation, cgroup setup, OverlayFS mount preparation) dominates. This pattern weakens on slower storage: HDD shows 15.3\% variation, macOS shows 21.7\%.

\textbf{Finding 2: Storage tier imposes a $2.04\times$ penalty.} Alpine warm-start on HDD (1157\,ms) is $2.04\times$ slower than SSD (568\,ms). This ratio increases with image complexity: nginx $2.28\times$, python $2.41\times$.

\begin{figure}[htbp]
\centering
\includegraphics[width=\columnwidth]{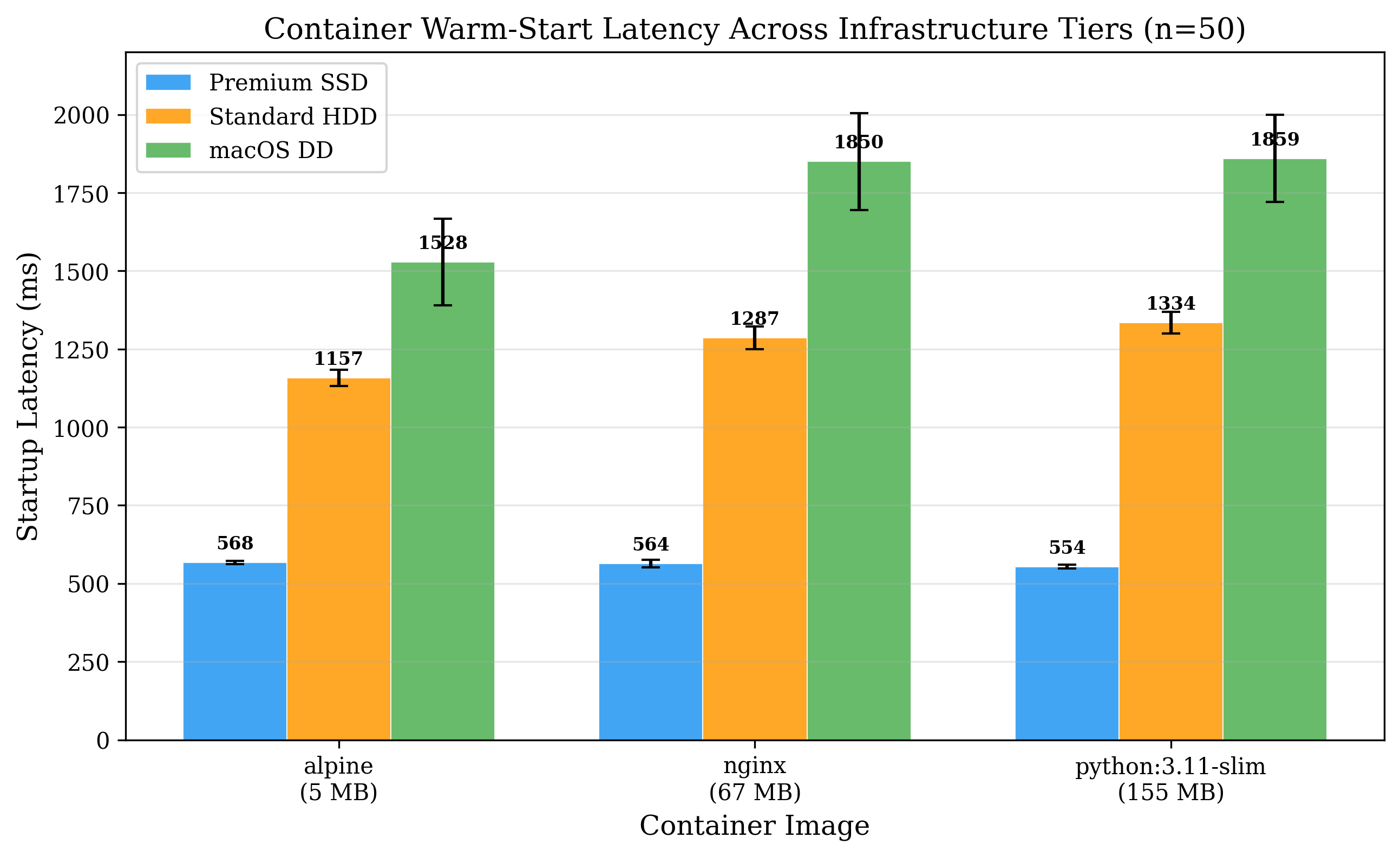}
\caption{Warm-start latency comparison across platforms and images. Error bars show 95\% confidence intervals. The $2.04\times$ HDD penalty and $2.69\times$ Docker Desktop penalty are visible across all image sizes.}
\label{fig:startup}
\end{figure}

\textbf{Finding 3: Docker Desktop virtualization tax is $2.69\times$.} macOS (1528\,ms) vs.\ SSD (568\,ms) for alpine. For nginx, the penalty reaches $3.28\times$. This originates from the LinuxKit VM, virtio block device, and hypervisor scheduling. Figure~\ref{fig:startup} visualizes this cross-platform comparison.

\subsubsection{Variance Analysis}

The coefficient of variation (CV) reveals scheduling jitter: Premium SSD CV$\!=\!3.4\%$, Standard HDD CV$\!=\!8.3\%$, macOS CV$\!=\!32.8\%$. The $9.6\times$ higher CV on macOS quantifies hypervisor-induced scheduling unpredictability.

\subsection{Platform-Invariant Primitives (RQ2)}

\subsubsection{Namespace Creation}

Namespace creation requires 7.94\,ms (SSD, $\sigma\!=\!2.05$) and 8.45\,ms (HDD, $\sigma\!=\!2.57$), contributing $<1.5\%$ of total startup. Near-identical results confirm this is CPU-bound, not I/O-bound---a platform-invariant primitive.

\subsubsection{CPU Throttling Accuracy}

\begin{figure}[htbp]
\centering
\includegraphics[width=\columnwidth]{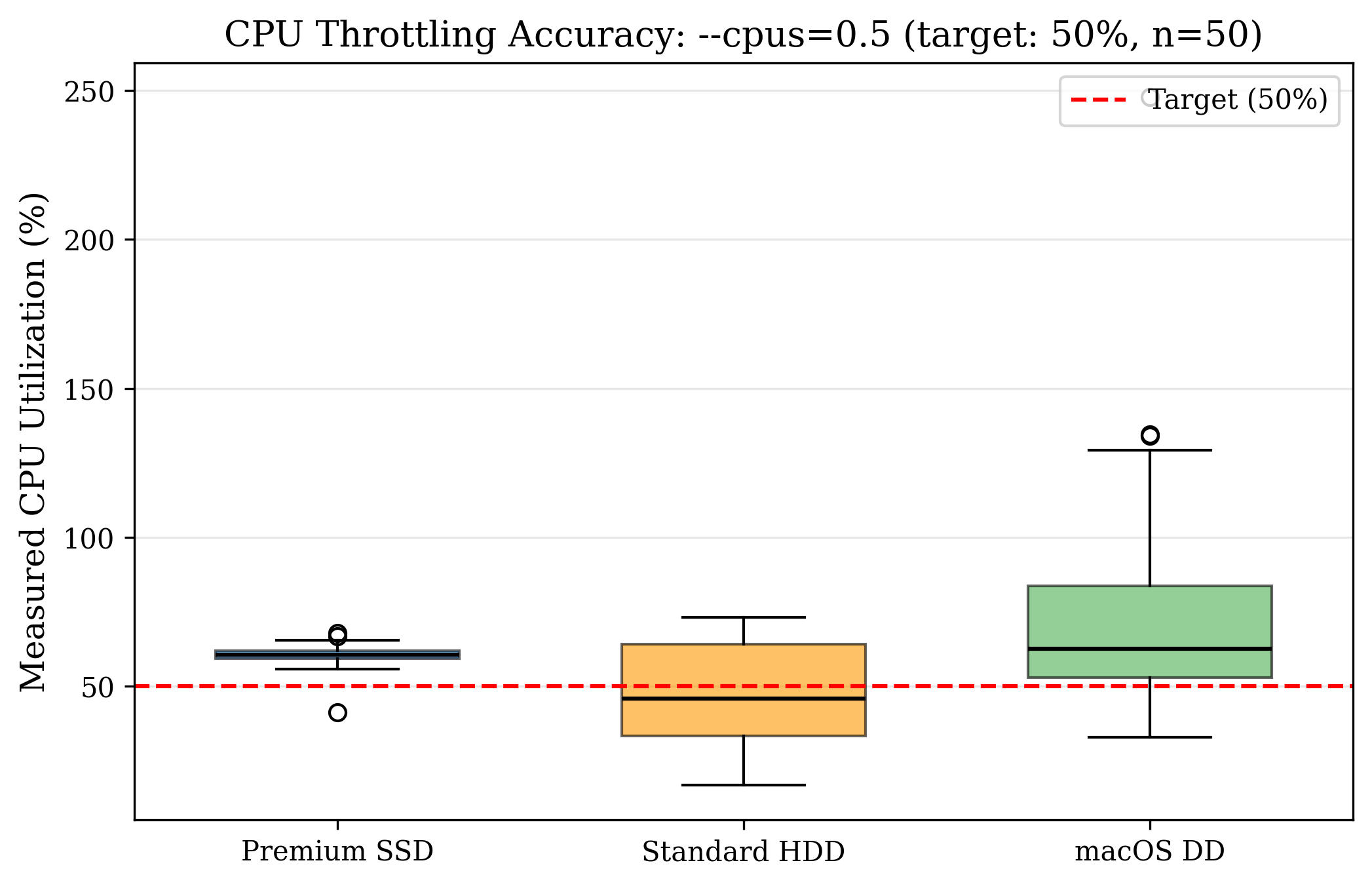}
\caption{CPU throttling accuracy under \texttt{-{}-cpus=0.5} (target: 50\%). Box plots show median, IQR, and outliers. macOS exhibits $9.5\times$ higher variance with outliers reaching 247\%, demonstrating compounded CFS scheduling inaccuracy through two-level virtualization.}
\label{fig:throttling}
\end{figure}

Under \texttt{-{}-cpus=0.5} (target: 50\%), Premium SSD achieves $\mu\!=\!60.52\%$ ($\sigma\!=\!3.79$), Standard HDD $\mu\!=\!48.21\%$ ($\sigma\!=\!16.24$), and macOS $\mu\!=\!71.63\%$ ($\sigma\!=\!36.14$). The $9.5\times$ variance increase from SSD to macOS demonstrates that two-level scheduling (macOS~$\to$~LinuxKit~$\to$~CFS) produces compounding inaccuracy. To our knowledge, this is the first empirical quantification of virtualization-compounded CFS scheduling variance. Figure~\ref{fig:throttling} illustrates the variance distribution across platforms.

\subsubsection{Network Overhead}

Bridge vs.\ host network overhead is 0.04\,ms on SSD and 0.06\,ms on HDD---negligible. On macOS, both modes show ${\sim}34$\,ms RTT ($17\times$ slower) because all traffic traverses the hypervisor's virtual network. The bridge-host difference (1.55\,ms) is not significant ($p\!=\!0.37$, Cliff's $\delta\!=\!0.104$).

\subsection{Storage Tier Impact (RQ3)}

\subsubsection{Sequential Write Performance}

Table~\ref{tab:writes} presents 256\,MB sequential write throughput. We report medians due to heavy right-skew in OverlayFS distributions (SSD OverlayFS median 1.10\,MB/s vs.\ mean 21.40\,MB/s). This skew arises from intermittent page-cache burst behavior: the Linux kernel occasionally satisfies the entire \texttt{dd} write from page cache before initiating disk flush, reporting completion at memory speed (${\sim}1$\,GB/s). In the majority runs, writes are synchronously committed to the storage device at the expected throughput. The bimodal pattern is consistent across both Azure tiers but absent on macOS, where the virtio layer enforces sequential commit ordering.

\begin{table}[htbp]
\centering
\caption{Write performance---median (MB/s), $n\!=\!50$.}
\label{tab:writes}
\small
\begin{tabular}{@{}lcccc@{}}
\toprule
\textbf{Platform} & \textbf{OverlayFS} & \textbf{Volume} & \textbf{Ratio} & \textbf{$p$} \\
\midrule
Prem.\ SSD & 1.1 & 194.0 & $0.006\times$ & --- \\
Std HDD & 1.3 & 136.6 & $0.010\times$ & --- \\
macOS DD & 238.1 & 224.5 & $1.06\times$ & 0.21 \\
\bottomrule
\end{tabular}
\end{table}

\begin{figure}[htbp]
\centering
\includegraphics[width=\columnwidth]{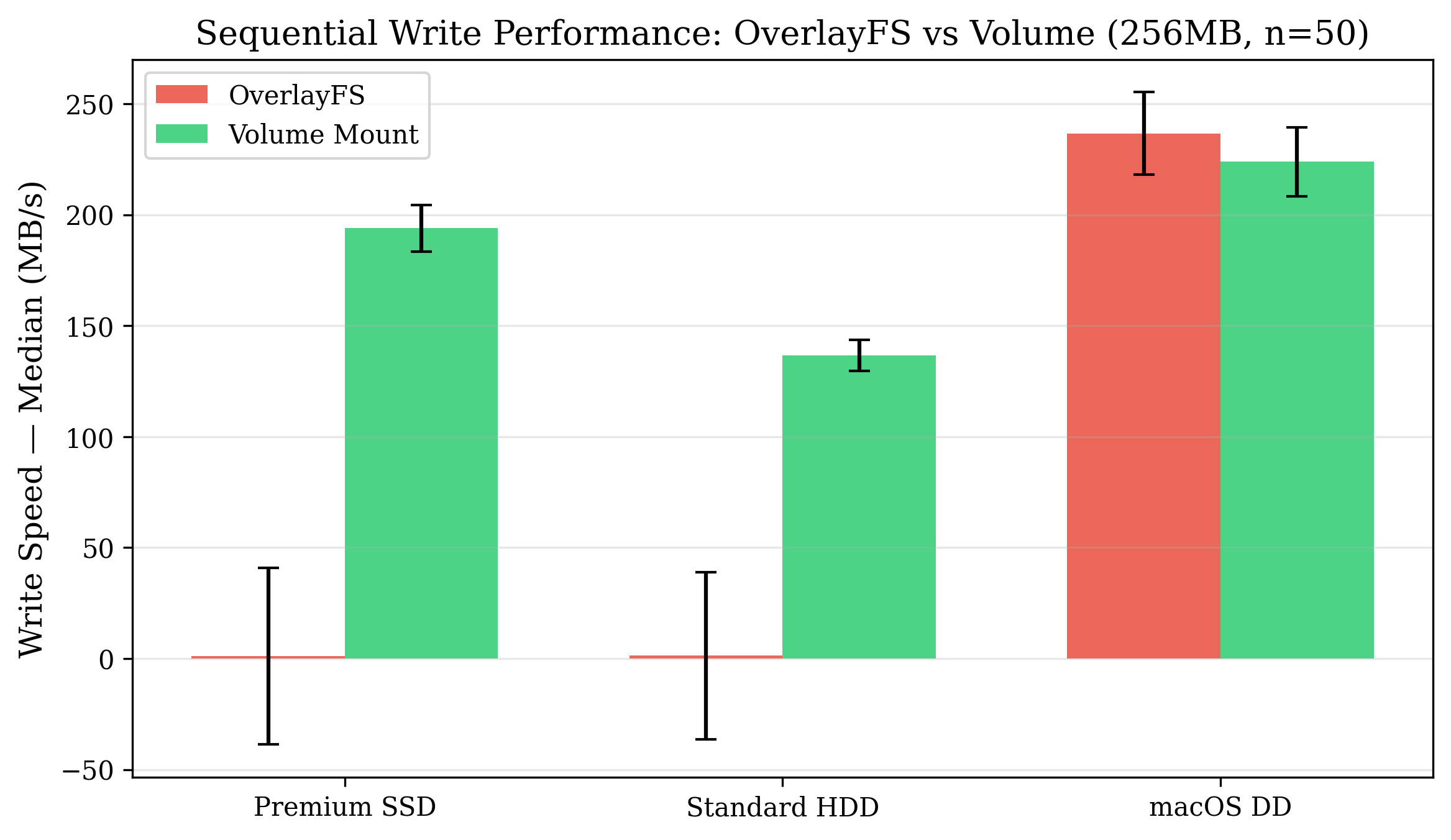}
\caption{Write throughput: OverlayFS vs.\ volume mounts. On Azure, OverlayFS collapses to $0.006\text{--}0.010\times$ of volume speed due to copy-up overhead. On macOS, the difference is not significant ($p\!=\!0.21$) because the virtio layer already serializes I/O.}
\label{fig:writes}
\end{figure}

On both Azure tiers, OverlayFS writes are two orders of magnitude slower than volume mounts because each write triggers copy-up, converting sequential I/O to random I/O. On macOS, performance is comparable ($1.06\times$, $p\!=\!0.21$, not significant), likely because Docker Desktop's virtio layer already serializes I/O. Figure~\ref{fig:writes} visualizes the OverlayFS write collapse.

\subsubsection{Metadata Operations}

Counterintuitively, OverlayFS is \emph{faster} than volume mounts for 500-file creation: on HDD, OverlayFS achieves 173\,ms vs.\ volume 834\,ms ($4.8\times$ faster). This occurs because OverlayFS metadata writes go directly to the upper directory without filesystem journaling overhead.

\subsubsection{Memory Efficiency}

Page cache sharing efficiency is approximately 0\% across all platforms: three identical nginx containers consume $3\times$ the memory of one (e.g., 9.88\,MB total vs.\ 3.29\,MB each on Azure). Each container's separate OverlayFS mount creates distinct page cache entries.

\section{Discussion}
\label{sec:discussion}

\subsection{Startup Decomposition Model}

Our measurements support a decomposition model:
\begin{equation}
T_{\text{startup}} = T_{\text{kernel}} + T_{\text{runtime}} + T_{\text{storage}}(\text{tier})
\end{equation}
where $T_{\text{kernel}} \approx 8$\,ms (invariant, $\sigma < 3$\,ms) and $T_{\text{storage}}$ varies $2.04\text{--}2.41\times$ between SSD and HDD. The dominance of $T_{\text{runtime}}$ over $T_{\text{kernel}}$ ($<1.5\%$ contribution) indicates that optimization targeting namespace or cgroup primitives yields negligible returns.

\subsection{OverlayFS Performance Paradox}

OverlayFS performs catastrophically for sequential writes ($0.006\text{--}0.010\times$ of volume speed) but outperforms volumes for metadata operations ($1.3\text{--}4.8\times$ faster). Sequential writes trigger copy-up (read-modify-write with random I/O); metadata operations write only to the upper directory, bypassing filesystem journaling. This inverts the conventional wisdom that volume mounts universally outperform OverlayFS.

\subsection{Implications for Practitioners}

\textbf{CI/CD pipelines:} The $2.04\times$ HDD-to-SSD ratio means a pipeline running 500 container operations per day saves ${\sim}5$ minutes by selecting SSD.

\textbf{Development environments:} The $2.69\times$ Docker Desktop penalty means performance benchmarks on macOS do not represent production behavior.

\textbf{Write-heavy workloads:} Applications performing sequential writes should use volume mounts. For metadata-heavy workloads (file creation, directory operations), OverlayFS actually outperforms volumes.

\textbf{CPU-sensitive workloads:} Docker Desktop's $9.5\times$ higher CFS variance makes it unsuitable for validating resource limit configurations.

\section{Limitations and Future Work}
\label{sec:limitations}

Our study evaluates a single VM size (2~vCPU, 8\,GB) and three container images; heavyweight images (e.g., CUDA containers) may exhibit different patterns. All measurements use sequential execution; concurrent scheduling may introduce contention effects~\cite{sharma2016}. On macOS, \texttt{drop\_caches} cannot reach Docker Desktop's VM caches, making cold-start measurements unreliable---the observed cold$\,<\,$warm inversion is an artifact. OverlayFS write performance shows extreme right-skew on Azure; we report medians to mitigate this.

Future work should extend to concurrent container scheduling, alternative runtimes (notably Podman, whose rootless mode uses user namespaces differently and may exhibit distinct namespace creation overhead), cgroups v1 vs.\ v2, additional storage tiers (NVMe, network-attached), and Kubernetes-orchestrated measurements.

\section{Conclusion}
\label{sec:conclusion}

This paper presented a systematic decomposition of Docker container startup performance across three infrastructure tiers with statistically rigorous measurements. Container startup decomposes into platform-invariant kernel operations (${\sim}8$\,ms, $<1.5\%$) and platform-dependent storage/runtime operations (550--1850\,ms). Image size contributes minimally (2.5\% variation on SSD). Storage tier imposes a $2.04\times$ penalty. Docker Desktop introduces a $2.69\times$ startup penalty and $9.5\times$ CFS variance increase. OverlayFS write speed is $0.006\text{--}0.010\times$ of volume mount speed, but metadata operations are $1.3\text{--}4.8\times$ faster on OverlayFS---inverting the conventional recommendation. These findings provide practitioners with empirical guidance for infrastructure selection and optimization prioritization.

\section*{Acknowledgments}
Benchmark infrastructure provisioned on Microsoft Azure. Docker Desktop testing on Apple Silicon. The author thanks the open-source community for the container ecosystem enabling this research.

\bibliographystyle{IEEEtran}

\end{document}